\magnification=1200
\font\titolo=cmbx10 at 14truept
\font\med=cmbx10 at 12truept
\parskip=3pt
\def\detg{\sqrt{|g|}}
\def\dettg{\sqrt{|\tilde g|}}
\def\sub#1{_{\scriptscriptstyle\rm #1}}
\def\section#1#2{\bigskip\goodbreak\item{\med #1}{\med #2}\medskip}
\def\dal{{\offinterlineskip\lower1pt\hbox{\kern2pt\vrule width0.8pt
\vbox to8pt{\hbox to6pt
{\leaders\hrule height0.8pt\hfill}\vfill
\hbox to6pt{\hrulefill}}\vrule \kern3pt}}}

\vbox to 8cm{\leftskip=0pt plus 1fill\rightskip=0pt plus 1fill\parindent=0pt

{\vskip0pt plus 3fil\titolo Are there metric theories of gravity other
than General Relativity?}
\vskip 0pt plus 1fil
{\med Guido Magnano}

{\sl Istituto di Fisica Matematica ``Joseph--Louis
Lagrange'',

via Carlo Alberto 10, I--10123 Torino, Italy}
\vskip 0pt plus 2fil
Talk given at the XI Italian Conference on General Relativity and Gravitation

Trieste, Sept 26--30, 1994
\vfil
}\vfill
\noindent {\bf Summary:} Current generalizations of the classical
Einstein--Hilbert Lagrangian formulation of General Relativity are
reviewed. Some alternative variational principles, based on different
choices of the gravitational field variable (metric tensor, affine
connection, or both) are known to reproduce -- more or less directly --
Einstein's gravitational equations, and should therefore be regarded as
equivalent descriptions of the same physical model, while other
variational principles (``Scalar--tensor theories'' and ``Higher--derivative
theories'') yield pictures of the gravitational interaction which
appear to be, {\it a priori\/}, physically distinct from GR. Such theories,
however, are also known to admit a reformulation (in a different set
of variables) which is formally identical to General Relativity
(with auxiliary fields having nonlinear self--interaction). The physical
significance of this change of variables has been questioned by several
authors in recent years. Here, we investigate to which extent purely affine,
metric--affine, scalar--tensor and purely metric theories can be regarded as
{\it physically\/} equivalent to GR. Focusing on the so--called ``nonlinear
theories
of gravity'' (NLG theories), which presently enjoy a renewed attention as
possible
models for inflationary cosmology, we show that if the metric tensor
occurring in a
nonlinear Lagrangian is identified by assumption with the physical spacetime
metric,
relevant physical properties (positivity of energy and stability of the vacuum)
cannot be assessed. On the other hand, using an alternative set of
variables (the
``Einstein frame'') one can prove that for a wide class of NLG theories
positivity
and stability properties do hold. This leads one to regard the rescaled metric
(Einstein frame) as the true physical one. As a direct consequence, the
physical
content of such ``alternative'' models is reset to coincide with General
Relativity,
and the ``Nonlinear  Gravity Theories'' become nothing but exotic
reformulations of
General Relativity in terms of unphysical variables.
\eject
\baselineskip=14pt
\section{I.}{Lagrangians for classical gravity}

Among the various classical theories of gravity which have been proposed
after the
formulation of General Relativity, it is not always easy to single out
those which
are truly ``alternative'' theories from those which turn out to be mere
reformulations of General Relativity itself. Even for the models which are
known to
be mathematically equivalent to GR, i.e., for which there exists a
well--defined rule
transforming the gravitational field equations into Einstein equations, the
problem
of the physical interpretation of such an equivalence often remains open.

We shall first comment on different cases of ``equivalence'' with
General Relativity. The considerations below, which are restricted to
classical (i.e.
non--quantum) aspects, refer to the four--dimensional framework, although
some of
the results quoted hold also in higher dimensions (the two--dimensional
case would
instead require a separate discussion).

In a relativistic theory of gravity, according to the common wisdom,
any configuration of the gravitational field should correspond to a unique
metric (and therefore causal) structure and to a unique geodesic structure on
spacetime. Hence, any (physical) solution of the field equations should
allow one to define on the spacetime manifold a metric tensorfield and an
affine connection. In the {\it purely metric} variational principles, only
the metric tensor acts as the dynamical variable, while in the
{\it metric--affine\/} Lagrangians the metric and the connection are both
present as
independent variables, their relationship being determined {\it a
posteriori\/} by the field equations. It is also possible to introduce {\it
purely affine\/} Lagrangians, which depend only on the connection. In other
models, now very popular, the gravitational interaction involves other
fields (typically, a scalar field) in addition to the ``geometrical'' degrees
of freedom. We are interested here in discussing to what extent the various
action principles lead to theories which physically differ from
General Relativity.

Two distinctive features of General Relativity, as far as
the gravitational field alone is considered, are the following:

\itemitem{(GR$_{1}$):} in the absence of matter, the metric tensor providing
the physical notion of spacetime distance obeys the vacuum Einstein equation;

\itemitem{(GR$_{2}$):} independently of external matter sources, the affine
connection which singles out the worldlines of free falling test particles is
the Levi--Civita connection of the physical metric tensor.

To determine completely the physical content of any theory of the
gravitational interaction one should, moreover, specify how gravity is
coupled to the  other fields. As a matter of fact, the various theories of
gravity are usually
classified according to their purely gravitational part, the determination of
matter coupling being viewed as a secondary problem. It is commonly believed
that a universal recipe to include matter interaction in General Relativity
is  provided by the ``minimal coupling'' prescription, but there is no
definitive evidence (neither theoretical nor experimental) that it should be
so (for instance, applying the minimal coupling prescription to the
Fierz--Pauli Lagrangian for a spin--2 field leads to inconsistencies
[1]). On the other hand, criteria to accept or reject possible
models of interaction are provided by energy conditions on the stress tensor
(see [2]).

The interaction of matter with gravity remains however crucial while
discussing the equivalence of different theories. We say that
two theories are {\it physically equivalent\/} iff

\item{(A)} the gravitational field has the same vacuum dynamics, and

\item{(B)} the coupling between the gravitational metric (or the
gravitational connection) and any kind of matter is the same for both
theories.

\noindent Let us now apply these criteria to the
most common types of gravitational Lagrangians.
\bigskip
\noindent{\bf PURELY AFFINE THEORIES} ---
The gravitational field variable is a
symmetric linear connection $\Gamma^{\alpha}_{\mu\nu}$. The first example of
a purely affine action principle has been introduced by Einstein and
Eddington:
$$
L\sub{EE}(\Gamma,\partial\Gamma)=\sqrt{|{\rm
det}R_{(\mu\nu)}|}\eqno{(1.1)}
$$
The metric structure associated to each solution is obtained using the
prescription:
$$
g^{\mu\nu}\detg={\partial L\sub{EE}\over\partial
R_{(\mu\nu)}}\eqno{(1.2)}
$$
(where $|g|$ denotes the absolute value of the {\it inverse\/} of the
determinant of
$g^{\mu\nu}$). This prescription is the prototype of a method that we shall
extensively use in the sequel; in dimension four and for the vacuum
Einstein--Eddington Lagrangian (1.1), the metric so defined turns out to be
proportional to the (symmetrized) Ricci tensor of
$\Gamma^{\alpha}_{\mu\nu}$. The second--order Euler--Lagrange equation
generated by
(1.1), upon insertion of the definition (1.2), becomes equivalent to
$$
\nabla_{\nu}(g^{\alpha\beta}\detg)=0\ ,\eqno{(1.3)}
$$
where $\nabla$ denotes the covariant derivative w.r. to $\Gamma$. Eq.~(1.3)
can be
satisfied only if
$\Gamma^{\alpha}_{\mu\nu}=\{^{\alpha}_{\mu\nu}\}_{_{\scriptstyle g}}$; hence,
the symmetric connection $\Gamma$ is forced by the field equations to
coincide with the Levi--Civita connection of $g$. By (1.2) $g_{\mu\nu}$ is then
necessarily proportional to its own Ricci tensor, and the solutions of the
vacuum
gravitational equations generated by the purely affine Lagrangian (1.1) are
the same
as the solutions of a vacuum Einstein equation with cosmological
constant. Hence, there is no doubt that the Einstein--Eddington
purely affine theory is physically equivalent to vacuum General Relativity.

The Einstein--Eddington Lagrangian (1.1) is the {\it only\/} covariant
scalar density
(of the appropriate weight) which can be built out of the symmetric part of
the Ricci tensor, up to a constant factor. Moreover, a
covariant Lagrangian cannot depend explicitly on the components
of the connection $\Gamma$, unless other fields are present. In this sense, the
Einstein--Eddington Lagrangian provides the only possible {\it vacuum\/}
purely affine model of gravity.

When matter interaction is implemented in this model, the situation becomes
more
complicated. We investigate this problem using a
Legendre--transformation technique, first introduced in this context by
Ferraris and
Kijowski [4].   Let
$$
L\sub{PA}=L\sub{PA}(R_{(\mu\nu)},\Gamma^{\alpha}_{\mu\nu},\Psi^{A},
\Psi_{\mu}^{A})\eqno{(1.4)}
$$
be any scalar density costructed using the (symmetrized) Ricci tensor of
$\Gamma$, $\Gamma$ itself, and other (unspecified) fields $\Psi$ with their
first derivatives $\Psi_{\mu}^{A}\equiv\partial_{\mu}\Psi^{A}$. One could
assume, for instance, that
$L\sub{PA}$ is the sum of the vacuum EE Lagrangian and some interaction
Lagrangian including covariant derivatives of $\Psi$ w.r.~to the connection
$\Gamma$. In particular cases (scalar field, electromagnetic field) only
ordinary derivatives will occur, and the Lagrangian will not depend on the
connection components $\Gamma^{\alpha}_{\mu\nu}$.

Let us introduce a ``conjugate
momentum'' to the connection, that we denote by $\pi^{\alpha\beta}$. The
{\it Legendre map\/}
relating $\pi^{\alpha\beta}$ to the ``configuration and velocity variables"
$(\Gamma,R)$ is defined as follows:
$$
\pi^{\mu\nu}={\partial L\sub{PA}\over\partial
R_{(\mu\nu)}}.\eqno{(1.5)}
$$
The Legendre transformation allows one to recast the original action principle
into a dynamically equivalent one, where the fields $(\Gamma,\pi)$ are
regarded as
independent variables. One should first compute the {\it inverse Legendre
map\/}
$r_{\mu\nu}=r_{\mu\nu}(\pi^{\alpha\beta},\Gamma^{\alpha}_{\mu\nu},\Psi^{A},
\Psi_{\mu}^{A})$, which is implicitly
defined by the following relation:
$$
{\partial L\sub{PA}\over\partial
R_{(\mu\nu)}}\Big|_{R_{(\mu\nu)}= r_{\mu\nu}(\pi,\ldots)}\equiv
\pi^{\mu\nu}\ .
\eqno{(1.6)}
$$
Then, one is able to introduce a new Lagrangian, the ``Helmholtz
Lagrangian'':
$$
L\sub{H} = \left[ R_{(\mu\nu)}-
r_{\mu\nu}\right]\pi^{\mu\nu}+L\sub{PA}(r_{\mu\nu},\Gamma^{\alpha}_{\mu\nu},
\Psi^{A},
\Psi_{\mu}^{A});
\eqno{(1.7)}
$$
the variation of $L\sub{H}$ w.r.~to the {\it independent\/} variables
$\Gamma$, $\pi$ and $\Psi$ is equivalent to the variation of
$L\sub{PA}$ w.r. to $\Gamma$ and $\Psi$ only. In fact,
$$\eqalign{
\delta L\sub{H} =& \left[R_{(\mu\nu)}- r_{\mu\nu} - {\partial
r_{\rho\sigma}\over\partial \pi^{\mu\nu}}\pi^{\rho\sigma}+
{\partial L\sub{PA}\over\partial r_{\rho\sigma}}{\partial
r_{\rho\sigma}\over\partial
\pi^{\mu\nu}}\right]\delta\pi^{\mu\nu}
+\cr &+\pi^{\mu\nu}\delta R_{(\mu\nu)}+\left[- {\partial
r_{\rho\sigma}\over\partial\Gamma^{\alpha}_{\mu\nu}}\pi^{\rho\sigma}+
{\partial L\sub{PA}\over\partial r_{\rho\sigma}}{\partial
r_{\rho\sigma}\over\partial
\Gamma^{\alpha}_{\mu\nu}}+{\partial L\sub{PA}\over\partial
\Gamma^{\alpha}_{\mu\nu}}\right]\delta\Gamma^{\alpha}_{\mu\nu}\cr &+
\left[-{\partial
r_{\rho\sigma}\over\partial \Psi^{A}}\pi^{\rho\sigma}+
{\partial L\sub{PA}\over \partial r_{\rho\sigma}}{\partial
r_{\rho\sigma}\over\partial
\Psi^{A}}+{\partial L\sub{PA}\over \partial
\Psi^{A}}\right]\delta\Psi^{A}+\cr &+\left[-{\partial
r_{\rho\sigma}\over\partial \Psi_{\mu}^{A}}\pi^{\rho\sigma}+
{\partial L\sub{PA}\over \partial r_{\rho\sigma}}{\partial
r_{\rho\sigma}\over\partial
\Psi_{\mu}^{A}}+{\partial L\sub{PA}\over \partial
\Psi_{\mu}^{A}}\right]\delta\Psi_{\mu}^{A}. }
$$
On account of (1.6), the coefficient of $\delta\pi$ vanishes iff
$
R_{(\mu\nu)}= r_{\mu\nu}
$: hence, the remaining part of the variation,
$${\partial
L\sub{PA}\over\partial R_{(\mu\nu)}}\delta
R_{(\mu\nu)}+{\partial L\sub{PA}\over\partial
\Gamma^{\alpha}_{\mu\nu}}\delta\Gamma^{\alpha}_{\mu\nu}+
{\partial L\sub{PA}\over \partial
\Psi^{A}}\delta\Psi^{A}+{\partial L\sub{PA}\over \partial
\Psi_{\mu}^{A}}\delta\Psi_{\mu}^{A},
$$
coincides with the total variation of the original purely
affine Lagrangian. On the other hand, using the well--known Palatini formula
for the variation of the Ricci tensor,
$$
\delta R_{(\mu\nu)} = \nabla_{\alpha}\delta\Gamma^{\alpha}_{\mu\nu}-
\nabla_{(\mu}\delta\Gamma^{\alpha}_{\nu)\alpha},\eqno{(1.8)}
$$
and subtracting a full divergence, one finds that the variation of $L\sub{H}$
w.r. to $\Gamma$ gives the equation
$$
\nabla_{\alpha}\pi^{\mu\nu}-\nabla_{\lambda}\pi^{\lambda(\mu}
\delta^{\nu)}_{\alpha}=
{\partial L\sub{PA}\over\partial
\Gamma^{\alpha}_{\mu\nu}} .
\eqno{(1.9)}
$$
Whenever the tensor density $\pi$ (which is symmetric by construction) is
non--degenerate, it can be re--expressed in terms of a symmetric
tensor: $\pi^{\mu\nu}=g^{\mu\nu}\detg$; writing the Legendre map in terms of
the new variable $g$, one gets exactly the Einstein--Eddington prescription
(1.2).  Equation (1.9) implies that the covariant derivative of the metric
tensor
$g$ w.r. to the connection $\Gamma$ vanishes identically if
$L\sub{PA}$ does not depend explicitly on $\Gamma^{\alpha}_{\mu\nu}$.
This holds for the vacuum theory and for particular types of matter coupling,
as mentioned above. In such cases, one has
$\Gamma^{\alpha}_{\mu\nu}=\{^{\alpha}_{\mu\nu}\}_{_{\scriptstyle g}}$ and
the model is fully equivalent to General Relativity. Otherwise, the
connection $\Gamma$ is not the metric connection. Then, if we assume that
$\Gamma$ is
the gravitational connection, i.e. determines the worldlines of free falling
particles, postulate GR$_2$ above is violated and the theory
is {\it not\/} physically equivalent to General Relativity.

If $\Gamma$ is not the metric connection, one could try to restore the
equivalence
with GR by assuming that GR$_2$ holds by definiton, i.e., the physical geodesic
structure is determined by $\{^{\alpha}_{\mu\nu}\}_{_{\scriptstyle g}}$, while
$\Gamma^{\alpha}_{\mu\nu}$ has a different physical interpretation. In that
case, what about postulate GR$_1$?  As we have seen, the dynamical equation
generated
by the variation of the Helmholtz Lagrangian with respect to a variation of the
metric
$g$ (through its associate tensor density $\pi$) is
$$
R_{\mu\nu}(\Gamma)=r_{\mu\nu}(\pi^{\alpha\beta},\Gamma^{\lambda}_{\alpha\beta},
\Psi^{A},\Psi_{\alpha}^{A}). \eqno{(1.10)}
$$
To recast it in a more familiar form, recall that the difference between
the Ricci tensor of $\Gamma$ and the Ricci tensor of $g$ can be expressed in
terms of the first and second covariant derivatives of $g$ with respect to
$\Gamma$; namely, one has
$$\eqalign{
R_{(\mu\nu)}(\Gamma)-R_{\mu\nu}(g)=&\ \nabla_{(\mu} Q^{\alpha}_{\nu)\alpha}
-\nabla_{\alpha} Q^{\alpha}_{\mu\nu} + Q^{\alpha}_{\beta(\mu}
Q^{\beta}_{\nu)\alpha} - Q^{\alpha}_{\mu\nu} Q^{\beta}_{\alpha\beta}\cr {\rm
with}\quad Q^{\alpha}_{\mu\nu} =
\Gamma^{\alpha}_{\mu\nu}-\{^{\alpha}_{\mu\nu}\}_{_{\scriptstyle g}}
=&\ {1\over2}
g^{\alpha\beta}\left(\nabla_{\mu}g_{\nu\beta}+\nabla_{\nu}g_{\beta\mu}-
\nabla_{\beta}g_{\mu\nu}\right).}\eqno{(1.11)}
$$
Equation (1.9) allows one to replace, after some manipulations, the covariant
derivatives $\nabla g$ with terms containing $\Gamma$ and the covariant
derivatives of the matterfields $\Psi$ with respect to the Levi--Civita
connection of $g$. The final result is that the dynamical equation for $g$
can indeed be put in Einstein form, with an ``effective stress--energy tensor''
$T_{\mu\nu}$ which is not of variational origin but still allows one to
obtain consistent conservation laws.

Let us summarize the results concerning purely affine models
as follows:

\item{(i)} the affine theory is fully equivalent to GR whenever the right-hand
side of (1.9) vanishes identically;

\item{(ii)} if the r.h.s.~of (1.9) does not vanish, and therefore
the connection $\Gamma$ is not the metric connection of $g$, the theory would
still be equivalent to GR provided the Levi--Civita connection of $g$ is
assumed
to be the gravitational connection, while the tensor $Q^{\alpha}_{\mu\nu}$
(1.11)
is regarded as an external field representing other than gravitation.

In the latter case, it should be
stressed that minimal coupling of matter fields with the connection
$\Gamma$ does not entail, in general, that the same matter fields are minimally
coupled to
$g$: on the contrary, such a coupling would determine not only a
non--standard interaction of matter with gravity, but also a
direct interaction between $\Psi$ and the tensorfield $Q$, which in general has
no reasonable physical interpretation. To construct a purely affine
model being physically equivalent to GR in the presence of matter, according
to the criterion (B) above, one should instead find a suitable
(possibly non--minimal) coupling between $\Psi$ and $\Gamma$ in the affine
Lagrangian (1.4) such that, after Legendre transformation, the
matter coupling with $g$ which is assumed to hold in General Relativity is
recovered. Such interaction terms can be explicitly computed, at least in
some cases, using the inverse Legendre transformation; we discuss below
an explicit example of the analogous situation occurring in the case of
nonlinear
metric theories.

\bigskip
\noindent{\bf METRIC--AFFINE THEORIES} ---
The difference between metric--affine and purely metric variational principles
for gravity consists in regarding the metric and the connection as mutually
independent variables while taking the variation of the action
integral\footnote
{$^1$}{Some authors seem to take for granted that independent
variations of metric and connection (according to what is improperly called
``Palatini method'') lead to the same gravitational equations which are
obtained in
the purely metric variational scheme. As it
has been pointed out in [5], this is true for Lagrangians which depend linearly
on the curvature scalar $R$, but is otherwise false.}.
Metric--affine models are quite popular, and the well--known
``Einstein--Cartan'' theory of gravity is based on a suitable
generalization of the metric--affine framework.

The equivalence between
a particular class of metric--affine action principles and GR has been
described in
full detail in [5]; in analogy with the purely affine case, it holds for
the vacuum theory but is broken by generic matter couplings. Consider the
``nonlinear
metric--affine Lagrangian''
$$
L\sub{MA}=f(R)\detg+L\sub{mat}(g,\Gamma,\Psi,\partial\Psi),
\eqno{(1.12)}
$$
where $R=R_{\mu\nu}(\Gamma)g^{\mu\nu}$ is the ``metric--affine curvature
scalar''
obtained by taking the trace of the Ricci tensor of $\Gamma$ with the
metric $g$.
It seems reasonable to assume that covariant derivatives are everywere
defined by the
connection $\Gamma$, and hence that $L\sub{mat}$ does not contain derivatives
of the metric $g$  (we shall see below, however, that this assumption has
strong physical implications).

Suppose first that there is no matter interaction at all,
$L\sub{mat}\equiv0$. Then, the Euler--Lagrange equations generated by (1.12),
after few manipulations,
become
$$ \left\lbrace\matrix{
f'(R)R_{(\mu \nu )}(\Gamma)-{1\over 2}f(R)g_{\mu\nu}=0\cr\cr
\nabla_{\alpha}(f'(R)\sqrt g \ g^{\mu \nu})=0\ ,\hfill}\right. \eqno{(1.13)}
$$
where $\nabla_{\alpha}$ still denotes the covariant derivative with respect to
$\Gamma$. Taking the trace of eq. (1.13a) one obtains (in dimension four)
$$
f'(R)R-2f(R)=0\ . \eqno{(1.14)}
$$
Now, (1.14) is an algebraic (or transcendental) equation for
$R$: assuming that the function $f$ is analytic, it can have only a discrete
set of roots $\{\rho_i\}_{_{i=1,2,\ldots}}$ (unless it is identically
satisfied, which happens in $d=4$ if $f(R)=R^2$). Any solution of
(1.13a) must then have constant metric--affine curvature:
$$
R = \rho_i \eqno{(1.15)}
$$
One substitutes this value of $R$ into (1.13b) and, provided $f'(\rho_i)\ne 0$,
the resulting equation is
$$
\nabla_{\alpha}(g^{\mu \nu}\sqrt g )=0\ ;\eqno{(1.16)}
$$
which is nothing but the metricity condition for $\Gamma$. Consequently, the
Ricci tensor of $\Gamma$ should coincide with the Ricci tensor of $g$, and the
system (1.13) finally becomes
$$\left\lbrace\matrix{
R_{\mu\nu}(g)-{1\over 4}\rho_ig_{\mu\nu}=0
\cr\cr
\Gamma^{\alpha}_{\mu\nu}=\lbrace^{\alpha}_{\mu\nu}\rbrace_g\ .\hfill} \right.
\eqno{(1.17)}
$$
The only dynamical equation left is thus the Einstein equation in the
vacuum, with
cosmological constant
$\Lambda_i={\rho_i\over4}$. The function $f(R)$ occurring in the
original nonlinear
Lagrangian (1.12) is only reflected in the spectrum of possible values
$\lbrace\Lambda_i\rbrace_{_{i=1,2,\ldots}}$ of the cosmological constant.

In other words, each solution of the system (1.13) is completely
represented by the
metric $g$ of an Einstein space; each solution corresponds to a definite root
$\rho_i$, so that one would actually observe the same value of the cosmological
constant at all points of space--time. The cosmological constant, however,
can be
different for different solutions of the {\it same\/} system of equations, in
contrast to the case of General Relativity. In this sense, a
single vacuum metric--affine Lagrangian (1.12) is equivalent to a whole set of
Einstein--Hilbert Lagrangians, spanned by the allowed values of
$\Lambda$ determined by the function $f(R)$ through (1.14).

We now investigate the consequences of matter coupling, using
the Legendre transformation. For Lagrangians depending on the Ricci tensor
only through the metric--affine curvature scalar, the conjugate momentum is
a scalar
density $\pi$, and the Legendre map is defined as follows\footnote{$^2$}{This
definition is based on a general mathematical approach described in [6].}:
$$
\pi = {\partial L\sub{MA}\over\partial R} = f'(R)\detg\ .\eqno{(1.18)}
$$
It is more convenient to express $\pi$ in terms of the associate scalar field
$p =
\pi |g|^{-1/2}$; with this definition, and provided
$$
{\partial^2 L\over\partial R^2} \ne 0 \qquad\Rightarrow\qquad f''(R) \ne 0 \
.\eqno{(1.19)}
$$
we can find a local inverse $r(p)$ of the Legendre map, which fulfills the
identity
$$
f'[r(p)]\equiv p\ .\eqno{(1.20)}
$$
The Helmholtz Lagrangian dynamically equivalent to (1.12) is then
$$
L_H(g,\Gamma,\partial\Gamma,p,\Psi,\partial\Psi)=p[R_{\mu\nu}(\Gamma)g^{\mu\
nu}-r(p)]\detg+
f[r(p)]\detg+L\sub{mat} \ .\eqno{(1.21)}
$$
The total variation of $L_H$ with respect to the four independent
fields $g$, $p$, $\Gamma$ and $\Psi$ yields the system
$$
\left\lbrace\matrix{
pR_{(\mu \nu )}-{1\over 2}g_{\mu\nu}[p(R-r)+f(r)]-[p-f'(r)]{\partial
r\over\partial
g^{\mu\nu}}=-{\partial L\sub{mat}\over\partial g^{\mu\nu}}\cr\cr
R-r-[p-f'(r)]{\partial
r\over\partial p} = 0 \hfill\cr\cr
\nabla_{\alpha}[p\sqrt g \ g^{\mu \nu}]-\nabla_{\lambda}[p\sqrt g \
g^{\lambda(\mu}_{\phantom{\alpha}}
\delta^{\nu)}_{\alpha}]=
{\partial L\sub{mat}\over\partial
\Gamma^{\alpha}_{\mu\nu}}\hfill\cr\cr
{\delta L\sub{mat}\over
\delta\Psi^{A}}=0\hfill
}\right.\eqno{(1.22)}
$$
On account of (1.20), the first two equations simplify to
$$
\left\lbrace\matrix{
p[R_{(\mu \nu )}-{1\over 2}g_{\mu\nu}R]+{1\over
2}g_{\mu\nu}[p\cdot r-f(r)]=-{\partial
L\sub{mat}\over\partial g^{\mu\nu}}\cr\cr
R=r(p)\hfill}\right.\eqno{(1.23)}
$$
Inserting (1.23b) into the trace of (1.23a) one finds on the
left--hand side a function of $p$ alone:
$$
p\cdot r- 2f(r)=-g^{\mu\nu}{\partial
L\sub{mat}\over\partial g^{\mu\nu}}\ . \eqno{(1.24)}
$$
If the trace of the "matter stress tensor" relative to $g$ (which coincides
with
the partial derivative ${\partial L\sub{mat}\over\partial g^{\mu\nu}}$, since
$L\sub{mat}$ does not contain derivatives of the metric) vanishes identically,
we recover the result already shown: equation (1.24) is in fact equivalent to
(1.14). Otherwise, the field equations do not force the scalar field
$p$ to be constant, but rather to be a prescribed function of the arguments of
$L\sub{mat}$. It is thus evident that the nonmetricity of
$\Gamma$, described by (1.22c), vanishes only if the matter Lagrangian is
independent of
$\Gamma$ {\it and\/} the stress tensor on the r.h.s.~of (1.22a) is traceless.

Thus, for particular kinds of matter fields (including the
relevant case of the electromagnetic field), the metric--affine nonlinear
Lagrangians (1.12) are directly equivalent to GR. For other types of matter
couplings,
$\Gamma^{\alpha}_{\mu\nu}\ne\lbrace^{\alpha}_{\mu\nu}\rbrace_{_{\scriptstyle
g}}$:
hence, to establish a physical equivalence one
should first assume that the Levi--Civita connection of $g$, rather than
$\Gamma$, should be regarded as the gravitational field. To recast (1.22a)
into a genuine Einstein equation for $g$, the difference between the Ricci
tensors of
$g$ and $\Gamma$ should then be included in a suitably defined ``effective
stress--energy tensor''; this would entail derivative couplings between the
various
fields and the tensor $Q^{\alpha}_{\mu\nu}=
\Gamma^{\alpha}_{\mu\nu}-\{^{\alpha}_{\mu\nu}\}_{_{\scriptstyle g}}$, which
would
be hardly acceptable on physical grounds.
Moreover, as for the purely affine case, one should be aware that
minimal coupling to $\Gamma$ would produce unphysical results, and minimal
coupling
to $g$ should be considered instead (a dependence of $L\sub{mat}$ on
$\{^{\alpha}_{\mu\nu}\}_{_{\scriptstyle g}}$ would contradict the assumption
made in (1.12), but would only marginally affect the computations).

As a side remark, let us suggest that the peculiar (and troublesome)
consequences of matter interaction in the nonlinear metric--affine framework
might yet offer some unexpected resources to the cosmologist.
According to our previous discussion, in a vacuum region of space--time the
physics
described by such models is the same as for General Relativity: $p$ is
constant and
determines the value of $\Lambda$ in that region. A layer of matter
separating two
vacuum regions may cause a transition between two different values of $p$,
and in that case one would observe different values of the cosmological
constant
in distinct vacuum regions of the same (connected) universe.

\bigskip
\noindent{\bf SCALAR--TENSOR THEORIES} ---
{}From the formal viewpoint, scalar--tensor models are purely metric
theories including a nonminimal coupling between a (positive--valued)
scalar field
and the curvature scalar $R$ of the metric $g$. Their prototype is the
(Jordan--Fierz--)Brans--Dicke Lagrangian:
$$
L\sub{BD}=\Big[\varphi R-
{\omega\over\varphi}g^{\mu\nu}\varphi_{,\mu}
\varphi_{,\nu}\Big]\detg.\eqno{(1.25)}
$$
The action can be generalized by allowing $\omega$ to depend on the scalar
field $\varphi$ and introducing a ``cosmological function'' $\lambda(\varphi)$
(see [7], [8]):
$$
L\sub{ST}=\Big[\varphi R-
{\omega(\varphi)\over\varphi}g^{\mu\nu}\varphi_{,\mu}
\varphi_{,\nu}+2\varphi\lambda(\varphi)\Big]\detg.\eqno{(1.26)}
$$
In such theories, however, the scalar field $\varphi$ is not supposed to
describe
gravitating matter, but rather an additional degree of freedom of the
gravitational
field. The gravitational field becomes thus a doublet consisting of a
spin--two field
(the metric $g$) and a spin--zero field: the
latter has no geometric significance but influences the coupling between
space--time
geometry and matter sources. From the physical viewpoint, therefore, the
Lagrangian
(1.26) should be regarded as a {\it vacuum\/} Lagrangian. The cosmological
function
seldom occurs in the current literature and in the sequel we neglect it,
assuming
$\lambda(\varphi)\equiv 0$.

Gravitating matter is represented by adding to (1.26) a standard
interaction Lagrangian, with minimal coupling to $g$. No direct coupling is
assumed
between matter and the spin--zero gravity field $\varphi$, since there is no
physical evidence at all for such an interaction. The full Lagrangian thus
becomes
$$
L=\Big[\varphi R-
{\omega(\varphi)\over\varphi}g^{\mu\nu}\varphi_{,\mu}
\varphi_{,\nu}+\ell\sub{mat}(\Psi,g)\Big]\detg.\eqno{(1.27)}
$$

A procedure known since 1962 as {\it Dicke transformation\/} [9] allows to
recast
a scalar--tensor Lagrangian into a standard Einstein--Hilbert one, by means of
a
conformal rescaling. One defines
a new metric
$$
\tilde g_{\mu\nu} = \varphi
g_{\mu\nu}; \eqno{(1.28)}
$$
in terms of the new
variables $(\tilde g_{\mu\nu},\varphi)$ the action becomes (up to a full
divergence term)
$$
L= \Big[\tilde R-
\left(\omega(\varphi)+{3\over 2}\right)\varphi^{-
2}\tilde g^{\mu\nu}\varphi_{,\mu}\varphi_{,\nu}
+\varphi^{-2}\ell\sub{mat}(\Psi,\varphi^{-1}\tilde g)\Big]\sqrt{|\tilde
g|},\eqno{(1.29)}
$$
($\tilde R$ being the curvature scalar of the metric $\tilde g$) and after a
redefinition  of the Brans--Dicke scalar,
$$ d\phi \equiv
\left(\omega(\varphi)+{3\over
2}\right)^{1\over2}{d\varphi\over\varphi}\ ,
\qquad\omega>-{3\over 2} \eqno{(1.30)}
$$
it takes the standard
form of the action for a linear massless scalar field minimally
coupled to the metric, at the price of introducing a coupling between
the external matter and the scalar field $\phi$:
$$
L= \Big[\tilde R -
\tilde g^{\mu\nu}\phi_{,\mu}\phi_{,\nu}+\tilde\ell\sub{mat}(\Psi,\phi,\tilde g)
\Big]\sqrt{|\tilde g|},\eqno{(1.31)}
$$
where
$\tilde\ell\sub{mat}=\varphi^{-2}\ell\sub{mat}(\Psi,\varphi^{-1}\tilde g)$,
with $\varphi$ replaced by the function of $\phi$ defined by (1.30).
The possible insertion of a cosmological function in the scalar--tensor
Lagrangian (1.27) would merely end up in a
potential term for the scalar field $\phi$.

According to a common terminology, the original metric
$g$ is referred to as the {\it Jordan frame\/} while the rescaled metric
$\tilde g$
is said to provide the {\it Einstein frame\/}. Apart from the use of the word
``frame'', which seems objectionable, this terminology is somehow misleading
as it suggests that the rescaled metric obeys the Einstein equation, while the
original one does not. As a matter of fact, also the dynamics of the Jordan
frame
metric $g$ can be represented by an Einstein equation. After some manipulations
described in full detail in [10], the field equations for the metric
and the Brans--Dicke scalar field can be rewritten as follows ($\varphi>0$
everywhere by assumption):
$$
R_{\mu\nu}-{1\over2}Rg_{\mu\nu}= {1\over\varphi}
(\nabla_{\!\mu}\nabla_{\!\nu}\varphi-
g_{\mu\nu}\dal\varphi)
+{\omega\over\varphi^2}(\varphi_{\!,\mu}\varphi_{\!,\nu}
-{1\over2} g_{\mu\nu}\varphi^{,\alpha}\varphi_{,\alpha})
+{1\over\varphi}T_{\mu\nu}(\Psi,g), \eqno{(1.32)}
$$
$$
\dal\varphi = {1\over
2\omega+3}\left(T_{\alpha}{}^{\alpha}-
{d\omega\over d\varphi}\varphi^{,\alpha}\varphi_{,\alpha}
\right),\qquad
{\rm with}\quad T_{\mu\nu}\equiv
 -{1\over \detg
}{\delta\left(\ell\sub{mat}\detg\right)\over\delta g^{\mu\nu}}.
\eqno{(1.33)}
$$
Let us compare these equations with the Einstein--frame field equations,
generated by
(1.31). For simplicity we restrict ourselves to the case of Brans--Dicke
theory,
i.e.~we set $\omega\equiv{\it const.}$ Having defined
$\gamma=\left(\omega+{3\over2}\right)^{-{1\over2}}$, the equations for
$\tilde g_{\mu\nu}$ and for the scalar field
$\phi={1\over\gamma}\ln\varphi$ become:
$$
\tilde R_{\mu\nu}-{1\over2}\tilde R\tilde g_{\mu\nu} = \phi_{,\mu}\phi_{,\nu}
-{1\over2}
\tilde g_{\mu\nu}\tilde g^{\alpha\beta}\phi_{,\alpha}\phi_{,\beta}
+ e^{-\gamma\phi}T_{\mu\nu}\ , \eqno{(1.34)}
$$
$$ \buildrel\sim\over\dal\phi
= {\gamma\over 2} e^{-
\gamma\phi}T_{\mu\nu}\tilde g^{\mu\nu}. \eqno{(1.35)}
$$
In the latter equations, $T_{\mu\nu}$ is still the stress tensor defined in
(1.33),
now depending on the fields $(\Psi,\phi,\tilde g)$ through the original
variables
$(\Psi,g)$; the stress tensor corresponding to $\tilde\ell\sub{mat}$ in
(1.31) is
instead $\tilde T_{\mu\nu}=e^{-\gamma\phi}T_{\mu\nu}$. The symbol
$\buildrel\sim\over\dal$ denotes the d'Alembert wave operator associated to
the new
metric $\tilde g$.

One learns by comparing (1.32) and (1.34) that the structural difference
between the
Jordan frame and the Einstein frame representation of the field dynamics
lies only
in the different properties of the total stress--energy tensor for matter
and scalar gravity.

The contribution of the Brans--Dicke
scalar to the effective stress--energy tensor in (1.32) has unphysical
features: it
is linear in the second derivatives of $\varphi$, and therefore it does not
fulfill
the Weak Energy Condition. For this reason, in the Jordan frame it is
impossible to
prove that the total ADM energy is bounded from below and the ground state
vacuum
solution is stable against matter perturbation. In fact, the only available
criterion for this purpose is the Positive Energy Theorem [11], which holds
provided
the stress tensor satisfies the Dominant Energy Condition.
Notice that the failure of the Dominant Energy Condition does not
necessarily entail
that the energy is unbounded fom below and no stable ground state exists: in
fact, the {\it vacuum\/} scalar--tensor Lagrangian (1.26), being equivalent to
(1.31), has the same stability properies as GR. The stability of the theory,
however, can be proved only after rescaling the metric from the Jordan to the
Einstein frame, since in the Jordan frame the total stress--energy tensor
is always
indefinite.

Hence, the equivalence between scalar--tensor models and General Relativity
holds in a sense which is quite different from what we have encountered in the
previous examples of purely affine or metric--affine gravity. In the original
variables, the model is equivalent to a general--relativistic scalar field
with an
{\it unphysical\/} coupling to the metric, while in the rescaled Einstein
frame the
additional degree of freedom is instead represented by a linear
massless scalar field, minimally coupled. The scalar field is assumed to be
positive everywhere for physical solutions, therefore for such solutions the
conformal rescaling can be always performed globally on space--time.

As long as the scalar field is {\it a priori\/} interpreted
as representing part of the gravitational interaction in
the scalar--tensor theory, the equivalence with GR should apparently be
regarded as
purely mathematical. The fact that physical spacetime distances are assumed
to be
measured by the Jordan frame metric, rather than by the Einstein
frame metric, is however the {\it only\/} real difference between the
Brans--Dicke model and General Relativity with an external scalar field, since
the scalar--tensor Lagrangian (1.27) is nothing but the ordinary
Einstein--Hilbert
Lagrangian (1.31) written in an alternative set of variables.

\bigskip
\noindent{\bf PURELY METRIC GRAVITY THEORIES} ---
In the most popular versions of gravity theory, the only independent variable
representing the gravitational field is the metric tensor. It is impossible to
construct a covariant Lagrangian containing only first derivatives of the
metric
(unless a fixed background connection is introduced [12]), therefore a purely
metric Lagrangian is necessarily of second order, and depends on the
Riemann tensor
components. Thus, a generic purely metric Lagrangian (different from the
Einstein--Hilbert Lagrangian) generates fourth--order equations: such models
are
thus called {\it higher--derivative gravity theories\/}. A typical
higher--derivative Lagrangian, including matter interaction, has the
following form:
$$
L\sub{PM}=f(g_{\mu\nu}, R^{\alpha}{}_{\beta\mu\nu}, \Psi^A, \nabla_{\mu}\Psi^A)
\detg, \eqno{(1.36)}
$$
where the covariant derivatives and the curvature tensor are those relative
to the
metric $g$. In most of the current literature, quadratic dependence of
the Lagrangian on the curvature tensor is generally assumed (possibly including
a linear term of the Einstein--Hilbert type), while higher powers
are seldom considered\footnote{$^3$}{Quite recently, some authors ([13],
[14]) have
investigated Lagrangians including derivatives of the Riemann tensor (in
particular,
scalar terms of the form $\dal^k R$), in view of possible generalizations of
the
order--lowering technique already known for fourth--order
gravity, but without evident physical motivation.}.

Among all higher--derivative models, a subclass deserves special consideration,
namely the Lagrangians depending on the derivatives of the metric only through
a
(polynomial or analytic) function of the
curvature scalar $R$. These are called {\it nonlinear gravitational
Lagrangians\/}, and are commonly written in the form
$$
L\sub{NL}=f(R)\detg + L\sub{mat}(g_{\mu\nu}, \Psi^A, \nabla_{\mu}\Psi^A)
 \eqno{(1.37)}
$$
(as we shall see in the sequel, however, a more general form should be
considered).
The analogy with (1.12) is evident, but the dynamics is quite different in
the purely
metric variational framework.
The gravitational field equation is in fact
$$ f'(R)R_{\mu\nu}-{1\over2} f(R)g_{\mu\nu}
-\nabla_{\mu}\nabla_{\nu}f'(R) +g_{\mu\nu}\dal
f'(R)=T_{\mu\nu}\ , \eqno{(1.38)}
$$
where, as usual,
$T_{\mu\nu}=-
{1\over\detg}{\delta L\sub{mat}\over\delta g^{\mu\nu}}
$.

It is well known that equation (1.38) can be formally recast into the
equations of General Relativity with a scalar field. Many authors describe the
relationship between the two theories purely in terms of a conformal
transformation
of the metric, as in the case of scalar--tensor theories
(the history of the introduction of the conformal rescaling and the appropriate
references can be found in [10], [15]). This is probably the
easiest way to obtain this result as far as only the field equations are
considered, but the comparison of the whole lagrangian structure of the two
theories
requires a deeper analysis.

Similarly to what has
been done above while dealing with the metric--affine
case, let us introduce a scalar conjugate momentum $p$, with the Legendre map
$$ p =
{1\over\detg}
{\partial L\sub{NL}\over\partial R}=f'(R)\ .\eqno{(1.39)}
$$
Also, let $r(p)$ be a function such that
$f'[R]\big|_{R=r(p)}\equiv p$; such a function (possibly not unique) exists if
$f''(R)\ne 0$.
The Helmholtz Lagrangian
equivalent
to (1.37) is
$$ L_{_H}=p[R(g)-r(p)]\detg+
f[r(p)]\detg + L\sub{mat}
.\eqno{(1.40)} $$
The difference between (1.40) and (1.21) is that the independent fields are now
$g$, $p$ and $\Psi$ only; the variation of te action corresponding to (1.40)
yields the equations
$$
\left\lbrace\matrix{
pR_{\mu \nu }-{1\over 2}g_{\mu\nu}[p(R-r)+f(r)]+g_{\mu\nu}\dal
p-\nabla_{\!\mu}\!\nabla_{\!\nu}p=T_{\mu\nu}\cr\cr  R(g) = r(p) \hfill\cr\cr
{\delta L\sub{mat}\over
\delta\Psi^{A}}=0 ,\hfill
}\right.\eqno{(1.41)}
$$
where the stress tensor $T_{\mu\nu}$ is the same as in (1.38). Due to the
second
derivatives of
$p$ occurring in (1.41), which were absent in the metric--affine case
(1.22), the
trace of the first equation does no longer
produce an algebraic equation for
$p$, but rather a differential equation which can be used to recast the
system in the
following form  (excluding the points where $p=0$):
$$\left\lbrace\matrix{
G_{\mu\nu} = p^{-
1}\nabla_{\mu}\nabla_{\nu}p
-{1\over6}\left\lbrace p^{-1}f[r(p)]+r(p)\right\rbrace
g_{\mu\nu}+p^{-1}T_{\mu\nu} \equiv\theta_{\mu\nu} \hfill\cr\cr
\dal p = {2\over3}f[r(p)]- {1\over3}
p\cdot r(p)\ ;\hfill\cr}\right.\eqno{(1.42)}
$$
It is instructive to compare (1.42) with the equations (1.32) of scalar--tensor
theory. In fact, the ``Legendre transform'' of the nonlinear Lagrangian (1.37)
is (formally) a particular case of scalar--tensor theory, and the resulting
field equations can be put in Einstein form, introducing the ``effective
stress--energy'' tensor $\theta_{\mu\nu}$ which is not a variational
derivative,
but is still covariantly conserved due to Bianchi identities. Let us stress
that
the Einstein equation (1.42), as well as (1.32), has been obtained {\it
without\/}
any conformal transformation.

Mathematically, each purely metric
(PM) nonlinear theory can be regarded as a metric--affine (MA) theory where
the {\it additional constraint\/}
$\Gamma^{\alpha}_{\mu\nu}=\lbrace^{\alpha}_{\mu\nu}\rbrace_{_{\scriptstyle
g}}$ has
been imposed while taking the variation of the action. From this viewpoint, the
relationship between purely metric and metric--affine nonlinear gravity
theories can
be described as follows:

\item{(i)} When the metricity constraint is present (PM theories), the
degrees of
freedom in the second--order picture of the dynamics include a
metric tensor and a scalar field (besides possible external matter fields). The
metric tensor always obeys Einstein equations, while the self--interaction
potential
of the scalar field depends on the choice of the function $f(R)$ in the
Lagrangian
(but does not depend on the matter Lagrangian).

\item{(ii)} If the metricity constraint is removed (MA theories) the scalar
field
is ``frozen down'': its dynamical equation is substituted by an algebraic
equation,
including a term depending on the presence of matter interaction (r.h.s.~of
(1.24)).
In the vacuum case, $p$ acts as a cosmological constant, and the nonmetricity
of
the connection vanishes as a consequence of the field
equations\footnote{$^4$}{Therefore, in the vacuum case each solution of the
field equations of the MA theory is also a solution of the corresponding PM
model:
this can be easily seen by imposing $p=$const. in the system (1.42). The
classical
solutions of a MA model are thus a {\it proper subset\/} of the (much
larger) set of classical solutions of the PM model with the same
$f(R)$.}. If matter is present, instead, the connection is not metric and
the field
$p$ is nonconstant; yet
$p$ does not represent an independent degree of freedom, since
it is completely determined by the matter distribution.

Let us mention the fact that the existence of a second--order picture of the
dynamics, which (at least formally) coincides with General Relativity, is
not only a
feature of the nonlinear metric theories (1.37), but holds for most
higher--derivative models (1.36). If the purely metric Lagrangian depends
on the full
Ricci tensor, as is the case for the most general quadratic Lagrangian in
dimension
four\footnote{$^5$}{For $d=4$, the Lagrangian density
$(R^2-4
R^{\mu\nu}R_{\mu\nu}+R^{\alpha}{}_{\beta\mu\nu}R_{\alpha}{}^{\beta\mu\nu})
\detg$ generates trivial field equations and can be freely added to any action
integral, allowing one to remove any quadratic term containing the Weyl tensor;
such terms play therefore an effective role only in higher--dimensional
theories.},
the conjugate momentum is a rank--two tensor, and the Legendre transform of the
theory is a sort of ``bimetric'' theory rather than a scalar--tensor model.
For a
complete discussion, we refer the reader to [6], [15], [16].

\section{II.}{Determination of the physical variables in nonlinear gravity
theories
and physical equivalence with General Relativity}

We shall now present a {\it physical\/} motivation to introduce a conformal
rescaling in the case of nonlinear metric gravity
theories. As we have just recalled, a GR--like formulation can be obtained
without any redefinition of the metric: for this purpose, it is enough to
isolate
the additional spin--0 degree of freedom due to the occurrence of nonlinear
second-order terms in the Lagrangian,
and encode it into an auxiliary scalar field by means of the Legendre
transformation.
However, while the dynamical terms for the metric in equation (1.42a) are
exactly
the standard ones, the source terms containing the field $p$ are substantially
different from those expected for a gravitating scalar field, according to
General
Relativity: the effective stress-energy tensor $\theta_{\mu\nu}$ is plagued
by the
same unphysical features as the r.h.s.~of (1.32). As we have already seen
for the
case of scalar--tensor theories, the Dominant Energy Condition can be
restored by a
redefinition of the field variables. A redefinition of the scalar field alone
is
useless to this purpose, while the following conformal rescaling of the
metric by
the field $p$ yields the correct result:
$$
\tilde g_{\mu\nu} = p g_{\mu\nu} .\eqno{(2.1)}
$$
It can be checked by direct computation that this conformal transformation is
the {\it only one\/} which deletes the linear second--order
term $\nabla_{\mu}\nabla_{\nu}p$.
To get exactly the standard coupling terms, one
redefines also the scalar field,
$p\mapsto\phi$, by
$$p= e^{{\scriptstyle \sqrt{2\over3}}\phi} ;\eqno{(2.2)}$$
both definitions make sense only if $p>0$, which is not always satisfied (for
quadratic Lagrangians including a linear Einstein--Hilbert term, $p$ is
everywhere
positive for solutions close to the Minkowski vacuum; for further details
we refer
the reader to [10]). The Lagrangian then becomes (up to a full divergence)
$$
\tilde L=\left[\tilde R-
\tilde g^{\mu\nu}\phi_{,\mu}\phi_{,\nu}-V(\phi)\right]
\dettg \eqno{(2.3)}
$$
with the potential
$$
V(\phi)= e^{-{\scriptstyle \sqrt{2\over3}}\phi}\,r[p(\phi)]-
e^{-2{\scriptstyle \sqrt{2\over3}}\phi}f(r[p(\phi)]) \eqno{(2.4)}
$$
In the ``rescaled'' Lagrangian (2.3), only the self--interaction potential
$V(\phi)$
keeps trace of the original nonlinear Lagrangian (1.37), while the
dynamical terms
are ``universal'' and are exactly those required by General Relativity.
Borrowing the
terminology from scalar--tensor theories, we say that the set of variables
$(g,p)$ provides the ``Jordan frame'', while the pair $(\tilde g,\phi)$
defines the
``Einstein frame'' for nonlinear theories.

Now, long--posed questions such as {\sl ``are nonlinear metric theories of
gravity
physically equivalent to General Relativity?''} or {\sl ``which
is the physical significance of the conformal rescaling?''} can be
phrased in rigorous and unambiguous terms. The problem is actually reduced
to the
following: {\sl``which metric is the true gravitational metric?''}
In fact, both metrics obey Einstein equations; therefore, according to our
initial
remarks, whether the theory is {\it physically\/} equivalent to General
Relativity
or not is a property which should be assessed by considering the coupling of
the
other fields with the gravitational field. Since the coupling with the metric
is
strongly affected by the conformal transformation, it is evident that an
external
field cannot be coupled in a physically satisfactory way to both metrics at
the same
time.

Previous attempts to determine the physical metric on theoretical grounds were
based on two distinct kinds of arguments: a first group of authors (e.g.
[17], [18])
tried to show that only one picture of the theory (i.e., either the Jordan
frame
or the Einstein frame) leads to a consistent formulation, while the other
entails a breakdown of the expected conservation laws. The second group of
authors
(see [10] for refs.) observed that the matter fields are minimally coupled
to the
original metric
$g$ in the Lagrangian (1.37), and regarding the rescaled metric as the
gravitational
field would produce unwanted effects such as nonconstant masses depending on
the
scalar field $\phi$.

The consistency argument, however, typically rests on calculations in which
the stress--energy tensor is computed by taking the variational derivative of
$L\sub{mat}$ with respect to one metric, while covariant
derivatives are taken using the Levi--Civita connection of the {\it
other\/} metric.
A careful analysis shows that such calculations are ill--grounded, and
appropriate
conservation laws do hold in both frames [10]. The argument based on matter
coupling,
on the other hand, is a sort of {\it petitio principii\/}. In fact, it is
evident
that the Lagrangian (1.37) is constructed by assuming {\it a priori\/} that
matter
should be minimally coupled to $g$. To deal with a concrete
example, suppose that one wishes to couple a (complex) charged scalar field
to higher--derivative gravity (with a quadratic Lagrangian) and to the
electromagnetic field. According to (1.37), one would be lead to define the
Lagrangian as follows:
$$
L\sub{NL} = \Big[aR^2 + R -
g^{\mu\nu}D_{\mu}\psi(D_{\nu}\psi)^*- m^2\psi\psi^*-
{1\over 8\pi}F_{\alpha\mu}F_{\beta\nu}g^{\mu\nu}
g^{\alpha\beta}\Big]\detg ,\eqno{(2.5)}
$$
where $D_{\mu}\psi\equiv\partial_{\mu}\psi-ieA_{\mu}\psi$. In the corresponding
Einstein frame Lagrangian, the coupling of $\psi$ with $\tilde g$ becomes
unphysical, and the mass of $\psi$ is rescaled by a nonconstant factor
depending on $\phi$. However, unless there are {\it independent\/}
motivations to
assume that the Lagrangian should necessarily be as in (2.5), one might
consider
instead the following Lagrangian:
$$
L\sub{NL} =
\left\lbrace{[R-
g^{\mu\nu}D_{\mu}\psi(D_{\nu}\psi)^*
+{1\over 2a}]^2\over
4m^2\psi\psi^*+{1\over
a}}-{1\over
8\pi}F_{\alpha\mu}F_{\beta\nu}g^{\mu\nu}g^{\alpha\beta}-
{1\over 4a}\right\rbrace\detg\ , \eqno{(2.6)}
$$
which, in spite of its exotic appearance, reduces to the same vacuum
Lagrangian $(aR^2+R)$ when the field $\psi$ is ``switched off''. The Einstein
frame Lagrangian corresponding to (2.6) turns out to be
$$
\tilde L = \Big[\tilde R-
\tilde g^{\mu\nu}\phi_{,\mu}\phi_{,\nu}-
{(e^{-{\scriptstyle
\sqrt{2\over3}}\phi}-1)^2\over  4ae^{-{\scriptstyle \sqrt{2\over3}}\phi}}-
\tilde g^{\mu\nu}D_{\mu}\psi(D_{\nu}\psi)^*- m^2\psi\psi^*-
{1\over 8\pi}F_{\alpha\mu}F_{\beta\nu}\tilde g^{\mu\nu}
\tilde g^{\alpha\beta}\Big]\dettg, \eqno{(2.7)}
$$
and in this case one would easily accept the idea that the physical metric is
$\tilde g$. From this example one learns that the coupling of a
given metric to matter fields is in fact {\it determined\/} by the
physical significance ascribed to it, i.e., by its relation to the
physical metric. Thus, the physical metric should be
singled out {\it a priori\/}, i.e., already in the vacuum theory.

One might ask whether
nonlinear Lagrangians such as (2.6), yielding minimal coupling with the
Einstein
frame metric, can be systematically produced. Let us warn the reader that
the naive
procedure consisting in taking a vacuum nonlinear Lagrangian, performing a
conformal
rescaling, then adding to the Einstein frame Lagrangian the appropriate minimal
interaction term and finally rescaling back the metric by the same conformal
factor, leads to incorrect results. The inverse rescaling has to be done in a
more subtle way, and once again the fundamental role played by the Legendre
transformation becomes apparent. In [10] the reader can find a
detailed description of the correct method.

It could seem at this point that the two frames are equally satisfactory: both
pictures are consistent, and both allow minimal coupling with external matter.
However, we have seen that a {\it vacuum\/} nonlinear Lagrangian is
equivalent to a
scalar--tensor theory, which in turn reduces to General Relativity with an {\it
unphysical\/} effective stress--energy tensor in the Jordan frame, or with a
well--behaved stress--energy tensor in the Einstein frame. In the Einstein
frame, the
Positive Energy Theorem can be applied and one can prove the existence of a
stable
ground state; thus, one knows that the vacuum theory is stable (for  suitable
choices of the function $f(R)$, see [10]; the stability of a
quadratic Lagrangian was first proved in [19]). What about the stability of the
interacting model?

Let us revert to the example of the two Lagrangians (2.5) and (2.6).
In both cases, the Positive Energy Theorem cannot be applied directly to the
fourth--order picture (to our present knowledge), and we should rely on
the second--order picture, introducing the scalar field $p$ (or $\phi$). For
the
Lagrangian (2.5), although the matter Lagrangian fulfills the appropriate
energy
condition, the total stress--energy tensor in the second--order picture has
indefinite signature, due to the unavoidable contribution of the scalar field
$p$. Thus, in the Jordan frame one loses all control on the positivity of
the total
energy and on the stability of the theory, as soon as ordinary matter is
coupled to
gravity. On the contrary, inserting the same matter Lagrangian
(with the rescaled metric) in the Einstein frame Lagrangian, as was done in
(2.7), is
perfectly safe, since the scalar field $\phi$  gives a positive
contribution to the
total stress--energy tensor.

The stability of the model in the presence of matter coupling with the
Jordan metric can indeed be checked, upon transformation to the Einstein
frame: since
the stress--energy tensor of external matter is simply rescaled by the
conformal
factor (which is assumed to be positive), it turns out that a standard
coupling in
the Jordan frame does not break the Dominant Energy Condition.
However, even in this case, the  conserved quantity which
attains a minimum at the stable vacuum, and thus can be physically
identified with
the total energy of the system, is the ADM energy defined in the Einstein
frame,
{\it not\/} in the Jordan frame. For this reason the Einstein frame is the most
natural candidate for the role of  physical frame in nonlinear gravity
theories. The
Jordan frame formulation may instead be regarded as a useful tool to
circumvent the
problem of the physical nature of the scalar field: a nonlinear Lagrangian,
which
does not contain any scalar field, can be introduced as the primitive
object of the
theory, then the scalar field is generated as a by--product of the
transformation to
physical variables. Such a mechanism could be compared to the standard
procedure of
``spontaneous symmetry breaking'' in gauge theories.

In conclusion, according to our analysis, the correct way to formulate a
nonlinear
metric theory as a viable model of gravity (including matter) consists in
assuming
that the Einstein metric is the gravitational metric, and therefore that
the theory
is {\it physically equivalent\/} to General Relativity. This suggests a
negative
answer to the question raised in the title of this talk.

\section{}{Acknowledgements}
This talk, partially supported by GNFM--CNR, is mostly based on a joint
work with
Leszek Soko\l owski [10] and on previous work done in collaboration with Mauro
Franca\-viglia, Marco Ferraris and Igor Volovich.

\section{}{References}
\vbox{
\frenchspacing\pretolerance=2000\parindent=25truept
\def\paper#1#2#3#4#5#6{\item{\hbox to 20truept{[#1]\hfill}}
{#2,} {\it #3} {\bf #4} (#5) #6}
\def\book#1#2#3#4#5{\item{\hbox to 20truept{[#1]\hfill}} {#2,} {\it #3}, #4
(#5)}

\paper{1}{C. Aragone, S. Deser}{Nuovo Cim.}{57B}{1980}{33}

\book{2}{S.W. Hawking, G.F.R. Ellis}{The Large Scale
Structure of Space--time}{Cambridge Univ. Press}{Cambridge  1973}

\book{3}{A.S. Eddington}{The Mathematical Theory of Relativity}
{Cambridge Univ. Press}{Cambridge 1924}

\paper{4}{M. Ferraris, J. Kijowski}{Gen. Rel. Grav.}{14}{1982}{165}

\paper{5}{M. Ferraris, M. Francaviglia, I. Volovich}{Class. Quantum Grav.}{11}
{1994}{1505}

\paper{6}{G. Magnano, M. Ferraris, M. Francaviglia}{J. Math.
Phys.}{31}{1990}{378}

\paper{7}{T. Damour, G. Esposito--Far\`ese}{Class. Quantum Grav.}
{9}{1992}{2093}

\book{8}{C.M. Will}{Theory and Experiment in Gravitational
Physics}{Cambridge Univ. Press}{Cambridge 1981}

\paper{9}{R.H. Dicke}{Phys. Rev.}{125}{1962}{2163}

\paper{10}{G. Magnano, L.M. Soko\l owski}{Phys. Rev.}{D50}{1994}{5039}

\book{11}{G.T. Horowitz}{{\rm in:} Asymptotic Behavior of Mass and Spacetime
Geometry, {\rm Lect. Notes in Physics Vol. 202, ed. F.
Flaherty}}{Springer}{Berlin
1984}

\paper{12}{M. Ferraris, M. Francaviglia}{Gen. Rel. Grav.}{22}
{1990}{965}

\paper{13}{D. Wands}{Class. Quantum Grav.}
{11}{1994}{269}

\paper{14}{F.J. de Urries, J. Julve}{preprint IMAFF}{95/36}{1995}{}

\paper{15}{G. Magnano, M. Ferraris, M. Francaviglia}{Class.
Quantum Grav.}{7}{1990}{557}

\paper{16}{J. Alonso, F. Barbero, J. Julve, A. Tiemblo}{Class. Quantum Grav.}
{11}{1994}{865}

\paper{17}{C.H. Brans}{Class. Quantum Grav.}{5}{1988}{L197}

\paper{18}{S. Cotsakis}{Phys. Rev.}{D47}{1993}{1437}

\paper{19}{A. Strominger}{Phys. Rev.}{D30}{1984}{2257}}

\bye